\documentclass{article}

\usepackage[a4paper, total={6in, 8in}]{geometry}

\usepackage{lipsum}
\usepackage{endnotes}
\usepackage{natbib}
\usepackage{graphicx}
\usepackage{moreverb,url}

\usepackage{lipsum}  
\usepackage{footmisc}
\usepackage{enumitem}
\usepackage{float}

\usepackage{amsmath}
\usepackage{amssymb}
\usepackage{authblk}
\usepackage{tabularx}
\usepackage{array}
\usepackage[table]{xcolor}

\usepackage{csquotes}

\usepackage[latin1]{inputenc}

\usepackage{musicography} 

\newcommand{\citeauthorpos}[1]{{\citeauthor{#1}'s}}

\newcommand{\citepos}[1]{{\citeauthorpos{#1}~\citeyearpar{#1}}}

\title{Evolving music theory for emerging musical languages \\
\ \\ 
\normalsize In Music 2025, Innovation in Music Conference \\
 20--22 June, 2025,  Bath Spa University, Bath, UK}

\author[1,2]{\normalsize{Emmanuel Deruty}}
\affil[1]{\small Sony Computer Science Laboratories, 6 rue Amyot, 75005 Paris, France}
\affil[2]{\small Department of Architecture, Design and Media Technology, Aalborg University, Aalborg, Denmark}

\begin{document}
\date{}  
\maketitle

\vspace{2cm}

\begin{abstract}

This chapter reconsiders the concept of pitch in contemporary popular music (CPM), particularly in electronic contexts where traditional assumptions may fail. Drawing on phenomenological and inductive methods, it argues that pitch is not an ontologically objective property but a perceptual construct shaped by listeners and conditions. Analyses of quasi--harmonic tones reveal that a single tone can convey multiple pitches, giving rise to tonal fission. The perception of pitch may also be multistable, varying for the same listener over time. In this framework, the tuning system may emerge from a tone's internal structure. A parallel with the coastline paradox supports a model of pitch grounded in perceptual variability, challenging inherited theoretical norms.

\end{abstract}


\newpage

\vspace{2cm}

\section*{Introduction}\label{sec:introduction}

The `Methods for Pitch Analysis in Contemporary Popular Music' series \citep{deruty2025vitalictemperament,deruty2025vitalicnonharmonic,deruty2025multiple,deruty2025primaal} explores alternative conceptualisations of pitch in contemporary popular music; this discussion summarises and extends those investigations.

\vspace{.2cm}

A foundational assumption in both psychoacoustics and Music Information Retrieval (M.I.R.) holds that harmonic complex tones in musical contexts convey a single pitch, typically corresponding to their fundamental frequency ($f_0$). This view underpins much of the theoretical discourse on music. Yet its universality is contested -- both by historical debate, such as the Ohm-Seebeck controversy \citep{turner1977ohm}, and by contemporary practice, including the use of multiphonics in acoustic \citep{fallowfield2019cello} and electronic contexts \citep{deruty2025vitalictemperament,deruty2025vitalicnonharmonic,deruty2025multiple,deruty2025primaal}. A single quasi-harmonic signal may give rise to multiple perceived pitches, with listeners potentially disagreeing on which are heard \citep{deruty2025multiple}. This final observation, in particular, raises a central question: is pitch an ontologically objective property of a tone, or a listener-dependent construct?

\vspace{.2cm}

This chapter focuses on pitch transmission in contexts where traditional models fall short -- particularly in contemporary popular music (CPM) involving electronic sound production. We adopt a phenomenological stance \citep{husserl1983ideas}, setting aside inherited assumptions, and draw on inductive reasoning in the spirit of \citet{bacon1620novum}, beginning with specific observations rather than fixed axioms.

\vspace{.2cm}

The chapter unfolds in five parts. We first outline key assumptions about pitch in Western classical music, presented as a set of `specifications'. Particular attention is given to two: Specification~5 (one tone conveys one pitch) and Specification~9 (two tones convey two pitches), which we show to be mutually incompatible in the general case. Building on this tension, and drawing on the concepts of tonal fusion \citep{dewitt1987tonal} and fission, we illustrate how pitch perception may vary across listeners and listening contexts.

\vspace{.2cm}
 
We then examine how tuning and scale in CPM may result from tone structure, and how ambiguous tones can support perceptual multistability -- allowing shifting interpretations even for the same listener. Finally, we draw a parallel with the coastline paradox \citep{mandelbrot1967long}, suggesting that pitch, like measured length, may resist naive objectivity.

\vspace{.2cm}

Together, these sections argue for a more flexible, perception-centered model of pitch -- one that better reflects contemporary musical practices and challenges inherited theoretical frameworks.


\newpage
\section{The Western classical music framework \\for pitch transmission}\label{sec:westernframework}

\begin{figure}[htbp]
  \centering
  \includegraphics[width=1\columnwidth]{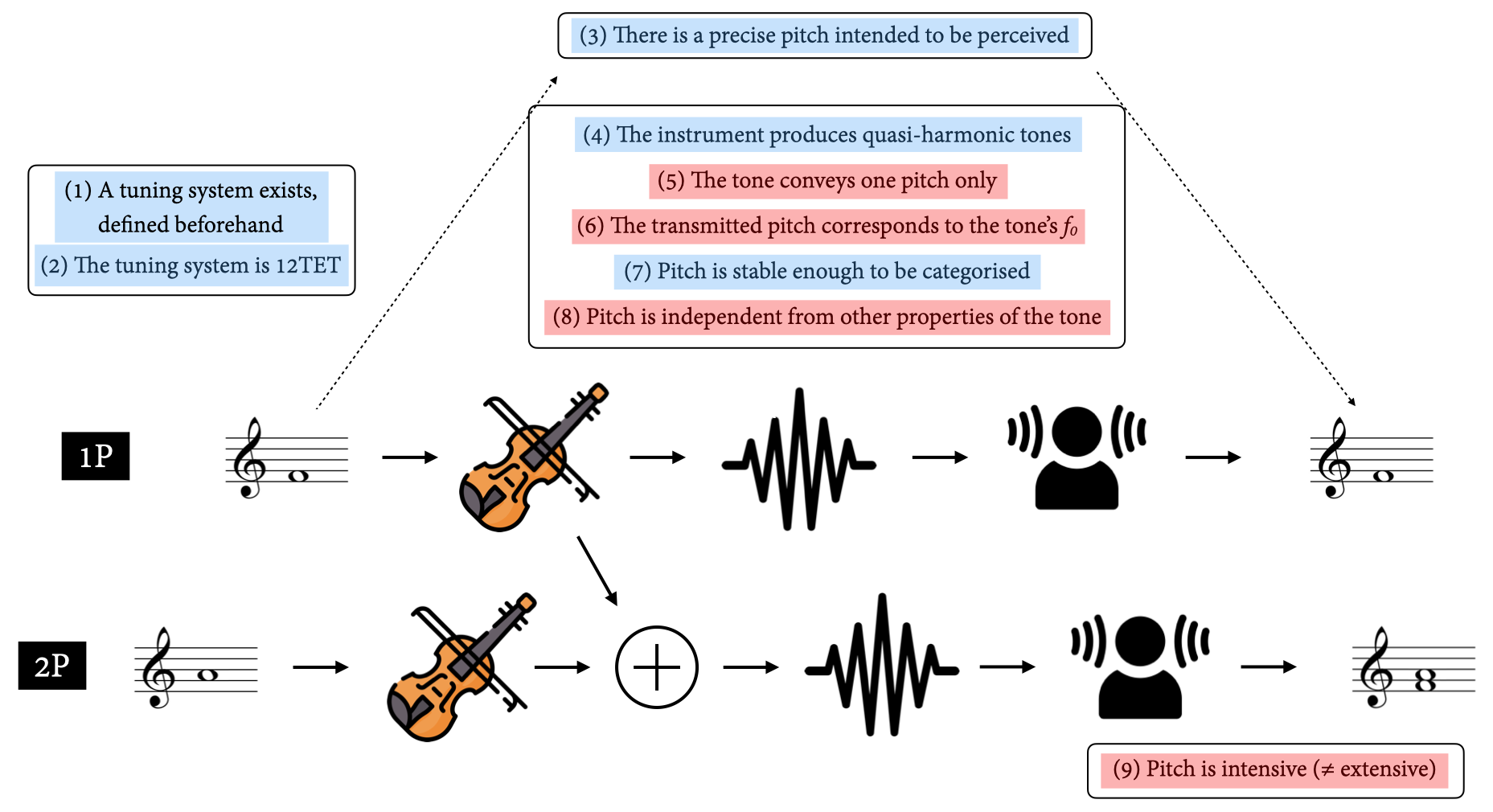}
  \caption{Specifications related to pitch transmission in Western classical music. `1P' denotes the transmission of one pitch; `2P' denotes the transmission of two pitches. The listed specifications fall into two categories, highlighted in blue (language conventions) and red (assumed properties of the tone). The leftmost frame gathers tuning system specifications, and the middle frame tone specifications.}
\label{fig:framework}
\end{figure}

In Western classical music, pitches are typically specified through musical notation. For each note written on the score, a performer uses an instrument to produce a tone intended to convey the corresponding pitch. The listener, in turn, is expected to perceive that pitch. When multiple notes are written to be played simultaneously, the listener is assumed to perceive each corresponding pitch. This framework is complemented by a number of specifications, illustrated in Figure~\ref{fig:framework}.

\vspace{.2cm}

The specifications in Figure~\ref{fig:framework} fall into two categories: conventions of the musical language (shown in blue) and assumed properties of the tone (shown in red). Music outside the Western classical canon -- including contemporary popular music (CPM), in the sense of \citet{deruty2022development} -- may follow different language conventions. In contrast, the properties of the tone are mostly assumed to hold across repertoires by both the Music Information Retrieval (MIR) and psychoacoustics literatures. The list below briefly describes each specification.

\begin{enumerate}

    \item \textbf{`A tuning system exists, defined beforehand'} is a language convention. Music can exist without a predefined tuning system -- as demonstrated in Ewe drumming traditions \citep{anku2009drumming}. Following \citet{deruty2025vitalictemperament,deruty2025primaal}, Section~\ref{sec:resulting} hypothesises that some forms of CPM may also operate without a fixed tuning system, with tuning instead emerging from the internal structure of the tones.

    \newpage

    \item \textbf{`The tuning system is 12-TET'} is a language convention. Although 12-TET is the current standard in Western music \citep{lindley2001equaltemperament}, alternative systems are employed in non-Western traditions -- such as Thai music \citep[p.~22]{morton1976traditional} -- and in early Western music as well \citep{lindley2001temperaments}. In relation to CPM, the point is mentioned in \citet{deruty2025primaal} and forms the core of \citet{deruty2025vitalictemperament}.

    \item \textbf{`There is a precise pitch intended to be perceived'} is a language convention. Music without clearly transmitted pitch exists in non-Western traditions, even when pitch-related phenomena are central to the musical language -- for instance, in Balinese gamelan music \citep[p.~6]{vitale2021balinese}. Certain strands of Western contemporary classical music likewise avoid specifying precise pitch values \citep{nyman1999experimental}. In the context of CPM, \citet{deruty2024pitchstrength} highlights various uses of weak pitch transmission as a stylistic feature, while \citet{deruty2025multiple,deruty2025primaal} show that pitch ambiguity plays a central role in production practices.

    \item \textbf{`The instrument produces quasi-harmonic tones'} is a language convention. Western classical instruments generally produce quasi-harmonic complex tones \citep{fletcher1993nonlinear}, but inharmonic instruments such as the gamelan or handpan are musically viable alternatives \citep{alon2015analysis}. In the context of CPM, most tones analysed in \citet{deruty2025vitalictemperament,deruty2025vitalicnonharmonic,deruty2025primaal,deruty2025multiple} are quasi-harmonic, with exceptions \citep{deruty2025vitalicnonharmonic}.

    \item \textbf{`The tone conveys one pitch only'} is an assumed property of the tone. Quasi-harmonic complex tones are generally recognised to convey a single pitch, both in the M.I.R. literature \citep{ewert2014score,gfeller2020spice} and in psychoacoustics \citep{yost2009pitch,oxenham2012pitch}. However, this property has historically been debated \citep{turner1977ohm}, and several exceptions are acknowledged \citep{shepard1964circularity}; \citep[pp.~1705--1706]{yost2009pitch}.

    \item \textbf{`The transmitted pitch corresponds to the tone's $f_0$'} is an assumed property of the tone. This assumption underlies much of M.I.R. \citep{salamon2012melody} and psychoacoustics \citep{yost2009pitch}. However, as shown in \citet{deruty2025vitalictemperament,deruty2025vitalicnonharmonic,deruty2025primaal,deruty2025multiple} and discussed in Section~\ref{sec:critical}, it may not hold universally.

    \item \textbf{`Pitch is stable enough to be categorised'} is a language convention. Western classical music relies on discrete pitch categories. However, when pitch evolves continuously, it may not align with fixed scale degrees -- as seen in Xenakis' music \citep[pp.~10--11]{harley2004xenakis}. In CPM, \citet{bittner2017pitch} show that vocal pitch is often better represented by pitch contours than by discrete values, a phenomenon especially evident in rap \citep{komaniecki2020vocal}. Continuous pitch trajectories have also been observed in non-vocal tracks \citep{deruty2025vitalictemperament,deruty2025multiple,deruty2025primaal}.

    \item \textbf{`Pitch is independent from other properties of the tone'} is an assumed property of the tone. Pitch is typically treated as independent of loudness and timbre \citep{wei2022review, asa2024timbre}, although this independence has been questioned \citep{melara1990interaction,krumhansl1992perceptual,marozeau2003dependency,deruty2025vitalicnonharmonic,deruty2025primaal}. Sections~\ref{subsec:prevalence} and~\ref{sec:resulting} illustrate this continuum: in quasi-harmonic tones, partial amplitude not only affects timbre but also influences perceived pitches and tuning.

    \item \textbf{`Pitch is intensive ($\neq$ extensive)'} is an assumed property of the tone. As \citet{gfeller2020spice} explain, pitch is considered intensive: combining sources with different pitches produces a chord, not a single fused tone -- unlike loudness, which increases additively with the number of sources. As seen in Section~\ref{sec:critical}, this principle conflicts with that of tonal fusion as defined by \citet{stumpf1890tonpsychologie,dewitt1987tonal}.

\end{enumerate}

Section~\ref{sec:critical} examines Specifications 5 and 9 in greater detail.


\section{Methodological foundations}\label{sec:methodological}

This chapter, along with \citet{deruty2025vitalictemperament,deruty2025vitalicnonharmonic,deruty2025primaal,deruty2025multiple}, challenges fundamental and widely accepted assumptions about music. The approach draws on phenomenological \textit{epoch\'e}, in which the observer brackets out all preconceived beliefs in order to examine phenomena as they are experienced, without presuppositions \citep[pp.~33, 59--61]{husserl1913german}. Such a process can support an inductive path, where the observation of particulars is followed by careful generalisation toward broader axioms \citep[Book~I, Aphorism~XIX]{bacon1620novum}.

\citet[p.~14]{descartes1637} emphasised that such generalisation should rely on exhaustive enumeration. However, when studying the relationship between audio signals and pitch, exhaustive collection is impractical. The standard M.I.R. strategy of compiling large datasets proves unsuitable in this context, as observations are manual. Feature engineering is likewise inappropriate at this stage, as it presupposes prior knowledge of what is to be observed, thereby contradicting the principle of phenomenological \emph{epoch\'e}. Generalisation must therefore proceed by alternative means.

In this context, discussions with producers provide valuable insights. For example, Luc Leroy of Hyper Music \citep{deruty2025primaal} notes that using synthesisers breaks the equivalence between $f_0$ and perceived pitch, suggesting that assumptions 5 and 6 may frequently fail in electronic music. Similarly, Vitalic \citep{deruty2025vitalictemperament,deruty2025vitalicnonharmonic,deruty2025multiple} reports using the number of perceived pitches per tone as a compositional parameter, reinforcing the idea that Specification 6 may often not apply.

Another approach may reside in substitution by analogy
\citep[Book II, Aphorism XLII]{bacon1620novum}, drawing parallels with phenomena such as tonal fusion and its effects on counterpoint (Section~\ref{subsec:fission}), the historical development of temperaments (Section~\ref{subsec:fission}), visually multistable equilibria (Section~\ref{sec:multistable}), and the coastline paradox (Section~\ref{sec:coastline}).


\section{Critical examination of Specifications 5 and 9}\label{sec:critical}

\subsection{Specification 5: one harmonic tone, one pitch}

Specification 5, which assumes that a harmonic complex tone conveys only one pitch, has a long debate history. Although the prevailing view today is that listeners typically perceive the fundamental as the sole pitch, this has not always been the consensus \citep[Prop.~XI, pp.~208--212]{mersenne1636harmonie}; \citep[pp.~13--14]{rameau1750demonstration}; \citep{ohm1843definition,seebeck1843definition}; \citep[pp.~58--63]{helmoltz1885sensations}; \citep{plomp1964ear, turner1977ohm, plomp2001intelligent}. While the ability to perceive more than one pitch from a harmonic complex tone is now generally acknowledged and has been demonstrated with artificial stimuli (Shepard, 1964; Yost, 2009), it is considered unusual in musical contexts (Plomp, 1976; Dixon Ward, 1970).

Specification 5 has significant implications. Because harmonic tones are typically associated with a single perceived pitch, this pitch has often been treated -- in \citepos{mari2023philosophical} terms -- as an \textit{ontologically objective property}, meaning it is assumed to exist independently of any observer. This view reflects a \textit{naive realist approach} to measurement, where the perceived pitch is taken as the `true' value simply waiting to be uncovered. It forms the basis for most monophonic pitch-tracking algorithms \citep{drugman2018traditional,kim2018crepe,riou2023pesto}. However, if perceived pitch can vary across listeners and listening conditions -- as demonstrated by \citet{deruty2025multiple} in the case of synthetic quasi-harmonic tones -- then the assumption of objectivity no longer holds. In the terms of \citet[p.~2]{poincare1906}, such a principle must instead be understood as a \textit{disguised convention} -- that is, a methodological convenience rather than a universal truth.

\subsection{Specification 9: two harmonic tones, two pitches}

Specification 9 -- the assumption that combining tones with different pitches results in a chord rather than a single fused tone -- is challenged by the concept of \textit{tonal fusion}, introduced by \citet{stumpf1890tonpsychologie} and extensively studied by \citet{dewitt1987tonal}. It refers to the perceptual phenomenon in which two or more simultaneous tones, particularly those forming intervals with simple ratios (e.g., octaves or fifths), are heard as a single, unified sound rather than as distinct pitches. 

\citet[p.~84]{dewitt1987tonal} revisited Stumpf's original findings and conducted their own experiments. While the authors consider their results strong enough to `elevate Stumpf's principle of fusion from conjecture to established fact', and therefore making Specification 9 a disguised convention, the outcomes are not entirely consistent: quantitative measures of fusion vary across experiments. These discrepancies may stem from differences in experimental protocols (e.g., percentage of `one tone' responses, mean reaction times) and in the acoustic properties of the stimuli (unspecified for Stumpf; triangle and square waves in later studies).

\citet[p.~75]{dewitt1987tonal} suggest that tonal fusion depends on both the number and relative intensity of harmonics in the signal, making it a property of the tones -- and by extension, of their sources. To date, tonal fusion has not been systematically investigated across a wide variety of tone types, a gap particularly relevant for electronic music. Whereas \citet[p.~83]{dewitt1987tonal} assume a gradual decay of harmonic intensity -- an assumption reflected in the harmonic tone model proposed by \citet[Sec.~2.4]{mauch2010approximate} -- \citet{deruty2025vitalictemperament,deruty2025vitalicnonharmonic,deruty2025primaal,deruty2025multiple} show that electronic tones often exhibit more diverse spectral configurations.



\subsection{Specification 5 $\land$ Specification 9}

Specifications 5 and 9 cannot hold simultaneously in the general case, creating a logical conflict. Indeed, summing two harmonic tones distant from one octave, S5 predicts tonal fusion (perception of one pitch), whereas S9 predicts the perception of two pitches.

\begin{figure}[h!]
  \centering
  \includegraphics[width=1\columnwidth]{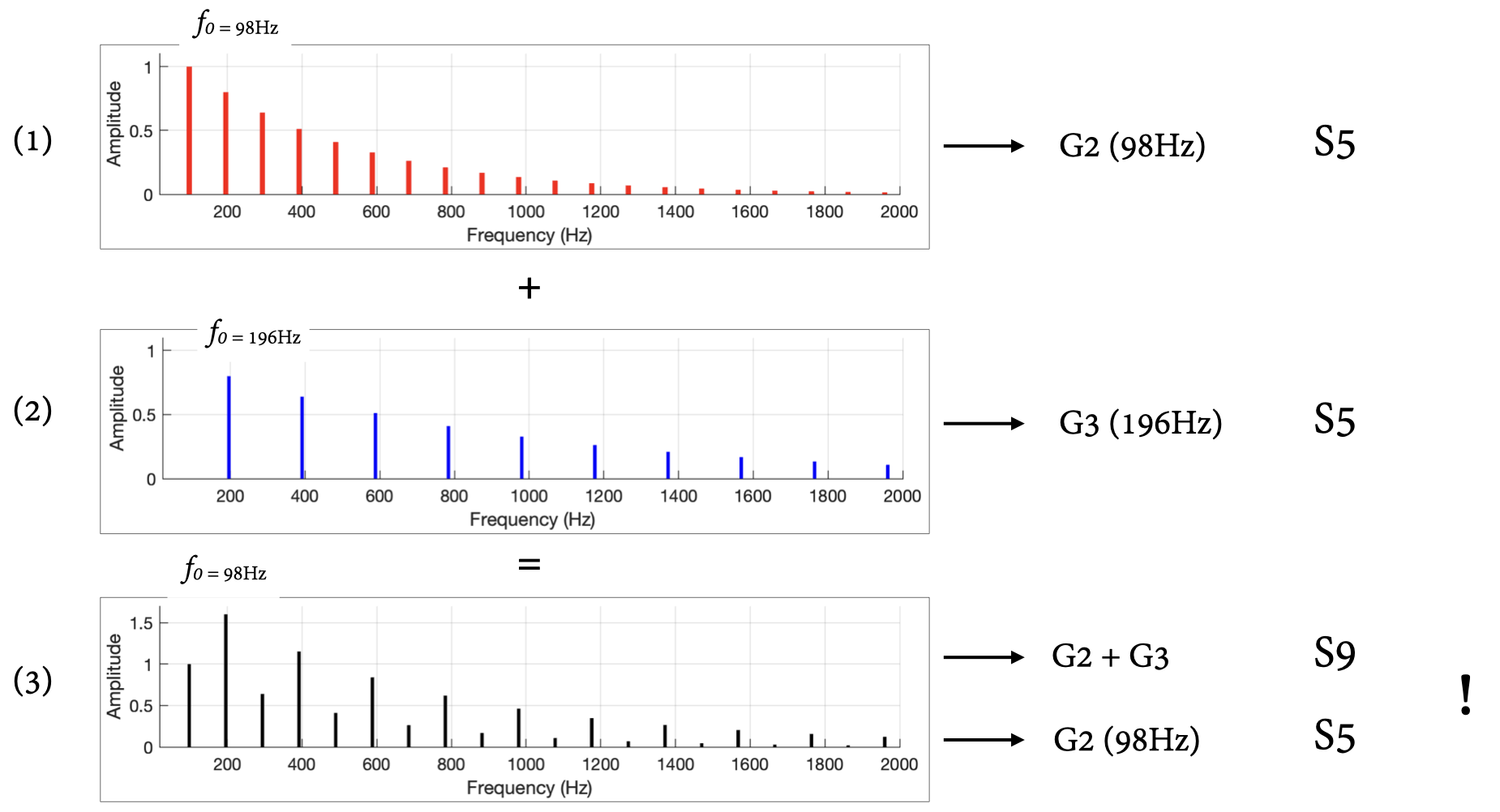}
  \caption{(1) Fourier transform (FT) of a harmonic complex tone with $f_0=98Hz$. (2) FT of a harmonic complex tone with $f_0=196Hz$. (3) FT of the sum of the two tones.}
\label{fig:S5S9}
\end{figure}

As shown in Figure~\ref{fig:S5S9}, consider a harmonic tone with $f_0=98Hz$ and partial energies following the model in \citet[Sec.~2.4]{mauch2010approximate} with $k = 0.8$. According to S5, the perceived pitch is G2. Now consider a second harmonic tone with $f_0=196Hz$ and the same spectral shape; under S5, the perceived pitch is G3. Summing the two tones yields a new harmonic complex tone with $f_0=98Hz$. According to S9, the perceived pitches should be G2 and G3. But if S5 is applied to the sum, the perceived pitch should be G2 only. Thus, at least in this case, S5 and S9 are incompatible.

Specifications 5 and 9 predict conflicting percepts for the sum of two tones. \citepos{dewitt1987tonal} experiments show that some listeners perceive a single pitch, consistent with S5, while others hear two distinct pitches, as predicted by S9. In other words, tonal fusion occurs only for some listeners. Although illustrated here with octaves (Figure~\ref{fig:S5S9}), similar effects are found with other intervals. This absence of perceptual consensus suggests that pitch is not an ontologically objective property of a harmonic tone.

\newpage
\subsection{Tonal fission}\label{subsec:fission}

\citet[p.~84]{dewitt1987tonal} conclude their study by suggesting that pitch perception ambiguities associated with tonal fusion could result from a single musical tone containing harmonics. This conjecture is supported by \citet{deruty2025vitalictemperament,deruty2025vitalicnonharmonic,deruty2025primaal,deruty2025multiple}, who describe the deliberate use of synthetic quasi-harmonic tones to convey multiple and ambiguous pitches. A similar phenomenon has previously been observed in multiphonics produced from acoustic sources \citep{fallowfield2019cello,fox2020art}.

\begin{figure}[h!]
  \centering
  \includegraphics[width=1\columnwidth]{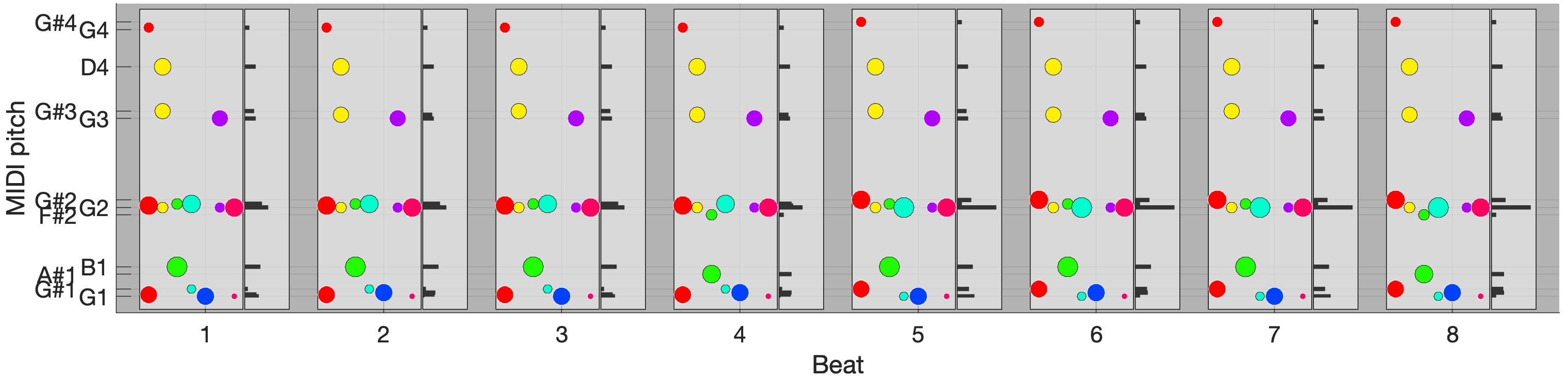}
  \caption{`Stamina' \citep{vitalic2012stamina}. Pitches perceived by seven expert listeners from a sequence of quasi-harmonic complex tones with an $f_0$ near G1. Each colour corresponds to a listener; dot area reflects the certainty of perception. Bar graphs show the aggregated pitch distribution across all listeners. Adapted from \citet{deruty2025multiple}.}
\label{fig:stamina}
\end{figure}

\citet{deruty2025multiple} report listening tests showing that (1) synthetic quasi-harmonic tones can produce multiple perceived pitches, most of them not corresponding to the $f_0$, and (2) perceived pitches vary across listeners and listening conditions. Figure~\ref{fig:stamina} presents transcriptions by seven expert listeners of a sequence of quasi-harmonic tones from a Vitalic bass line. Similar results were found for widely used harmonic tones such as guitar power chords and TR-808-style bass drums. Pitches with low listener agreement may be interpreted as having low pitch strength, as mentioned in \citet[Sec.~6]{deruty2024pitchstrength}.

The term \textit{fission} has been used to describe the perception of sounds as originating from multiple sources \citep{vannoorden1975coherence,bregman1994auditory,moore2012properties}. The tones from Vitalic, Hyper Music, the guitar power chord (a harmonic tone), and TR-808-style bass drums exhibit \textit{tonal fission}: the emergence of multiple perceived pitches from a single tone. Tonal fission -- where the number of perceived pitches exceeds the number of harmonic tones -- can be contrasted with tonal fusion, where fewer pitches are perceived than tones present. In both cases -- fusion and fission -- the perceived pitches appear to depend on the listener.

\subsection{Discussion: prevalence of tonal fusion}\label{subsec:prevalence}

Tonal fusion introduces ambiguity into pitch perception, supporting the hypothesis that pitch may not be an ontologically objective property of a tone. This challenges a core assumption in much of the M.I.R. literature, which often treats pitch as a fixed, measurable feature. While this chapter does not aim to present new quantitative data on tonal fusion, several lines of argument underscore its significance within Western music traditions.

A strong indication lies in \citepos{huron2001tone} association of tonal fusion with two well-known counterpoint rules in Western classical music -- the `Parallel Unisons, Fifths, and Octaves Rule' and the `Consecutive Unisons, Fifths, and Octaves Rule' -- both formalised by \citet{zarlino1558harmoniche}. According to Huron, these rules aim to maintain the perceptual clarity of individual musical voices by avoiding tonal fusion. \citet{huron2001tone} also observes that J.S. Bach systematically avoided isolated octaves and fifths for the same reason. Conversely, later works such as those of \citet{Chopin1835Etude} and \citet{Debussy1905Pelleas} made frequent use of these intervals, leading \citet{huron2001tone} to conclude that the desirability of tonal fusion is context-dependent, varying with the composer's perceptual and musical goals. In both cases -- whether prohibition or deliberate use -- if tonal fusion has shaped compositional practices across different stylistic periods, it is unlikely to be a marginal phenomenon.


\section{Resulting tuning system and scales}\label{sec:resulting}

This section questions Specification 1 in the case of CPM involving electronic means of production. In Western classical music, the term scale is generally associated with the coarse organisation of pitches \citep{drabkin2001scale}, while temperament and tuning refer to finer pitch adjustments \citep{lindley2001temperaments}.

When multiple pitches are perceived from a quasi-harmonic tone, at least some of these pitches derive not from the $f_0$ but from its partials. The precise positions of the partials determine tuning. Strictly harmonic tones would result in just intonation. However, \citet{deruty2025vitalictemperament,deruty2025vitalicnonharmonic,deruty2025primaal,deruty2025multiple}, and \citet{deruty2025vitalicnonharmonic} in particular, show that the studied tones are consistently slightly inharmonic -- though not enough, unlike the tones analysed by \citet{moore1986thresholds}, for inharmonicity alone to account for the perception of multiple pitches.

\newpage

In the compositional approaches of Vitalic \citep{deruty2025vitalictemperament,deruty2025vitalicnonharmonic,deruty2025multiple} and Hyper Music \citep{,deruty2025primaal}, the tone often precedes pitch organisation. Coarse-pitch positions are selected from available partials or partial groups, while fine-pitch values are retained as generated. As a result, both tuning and, to some extent, the scale emerge from the internal structure of the synthesised tone. Beyond \citet{deruty2025vitalictemperament,deruty2025vitalicnonharmonic,deruty2025primaal,deruty2025multiple}, see \citet[2'48]{musicradartech2014vitalicITV} for an interview in which Vitalic illustrates this approach.

The historical shift from modal polyphonic writing to harmonic and tonal writing has been described as a move from \textit{resultant harmony} to \textit{active harmony} \citep[p.~82]{coeurdevey1998histoire}. As noted by \citet[Book II, Chap. 19]{rameau1722traite} and detailed by \citet[p.~72]{dahlhaus1990studies}, in polyphony, harmony arises from the interaction of melodic lines -- the functions of compound sonorities being secondary to melodic structure. In harmonic writing, as formalised by \citet{caccini1978nuove}, harmony plays an active role: chords are no longer emergent but are perceived as independent, intentional unities.

Electronic music production may foster a comparable shift: pitches, rather than serving as foundational building blocks, may emerge from the tone's spectral structure -- making the tone itself the active element.

In music, \textit{idiom} typically refers to the effective use of an instrument's distinctive resources \citep[p.~103]{huron2009characterizing}. A musical goal is idiomatic to an instrument if it can be achieved with relative ease (p.~115). This concept parallels \citepos{jorda2005digital} notion of \textit{instrument efficiency}, defined as the ratio of musical output complexity to input complexity, adjusted for control diversity \citep[p.~54]{mcpherson2020idiomatic}. Because synthesisers allow precise control over partials, the use of active tones and resultant pitches can be considered an idiomatic compositional strategy for these instruments.


\vspace{.2cm}
\section{Multi-stable equilibria}\label{sec:multistable}

This section builds on observations from \citet{deruty2025multiple} to hypothesise that, in cases where no perceptual consensus exists regarding pitch, a single listener's perception of a given tone may also shift over time, resulting in a multistable equilibrium.

\textit{Multistability} occurs when a single physical stimulus leads to alternation between distinct subjective percepts \citep{schwartz2012multistability}. It was first described in vision, with the classic example being the Necker cube \citep[p.~336]{necker1832lxi}: a two-dimensional drawing of a cube lacking orientation cues, which can be seen either with the lower-left or upper-right square as the front face. In audio, examples of multistability (specifically bistability) include: (1) the tritone paradox, in which a pair of Shepard tones separated by a tritone can be perceived as either ascending or descending \citep{shepard1964circularity,deutsch1991tritone}; and (2) an alternating A-B tone sequence, which may be heard as a single stream (ABA-ABA) or as two distinct streams -- A-A-A-A and -B-B-B- \citep{pressnitzer2006temporal}.

\citet{deruty2025multiple} report that listeners confronted with harmonic complex tones producing multiple pitch percepts can sometimes deliberately choose which set of pitches to perceive. This suggests that quasi-harmonic complex tones may give rise to multistable percepts.

\newpage

One hypothesis would be that the deliberate construction of tones as multistable may help sustain listener interest. \citet{madison2017repeated} found that listener appreciation decreases when musical excerpts are repeated multiple times within a session. A similar dynamic appears in tracks by Vitalic and Hyper Music, which often involve heavy repetition. \citet{orbach1963reversibility} explain perceptual switching in the Necker cube through `satiation of orientation': the cognitive process favouring one interpretation becomes temporarily exhausted, prompting a shift to an alternative. This raises the hypothesis that, in multistable tones, producers may intentionally enable perceptual switching, delaying the onset of satiation and helping the music retain its appeal over repeated listens.

As a result, perception may actively structure the perception of pitch rather than passively receiving it. This hypothesis merits further investigation in future studies.


\vspace{.2cm}

\section{Parallel with the coastline paradox}\label{sec:coastline}

The ambiguity of pitch perception in harmonic tones shares features with the coastline paradox \citep{steinhaus1954length,richardson1961problem,mandelbrot1967long}. This section considers how in both cases: (1) an apparently straightforward concept fails under closer examination; (2) no ontologically objective property emerges through measurement; and (3) a quantity is observed to vary over $\mathbb{R}$, even though discrete values in $\mathbb{N}$ were initially expected. Additionally, both problems are well-suited to a resolution using an inductive process.

Although it may seem straightforward to ask, `What is the length of the coastline of Britain?', the answer depends on the ruler's length: shorter rulers capture more detail, causing the measured length to grow as ruler size decreases \citep{weisstein2025coastline}. Consequently, the length becomes infinite or, more precisely, undefinable \citep{mandelbrot1967long}. Plotting ruler length against measured coastline length on a log-log scale yields a straight line, whose slope corresponds to the fractal dimension -- a value between 1 and 2 \citep{weisstein2025coastline}. For example, the fractal dimension of West Great Britain's coastline has been estimated at approximately 1.25 \citep[pp.~25-33]{mandelbrot1983fractal}. As a result:

\begin{enumerate}[noitemsep,label={(\arabic*)}]
    \item The notion that a unique length is an objective feature of a line, though seemingly obvious, does not withstand closer examination.
    \item The length of a line depends on the observer's perspective, specifically the observation scale.
    \item The dimension of a line, previously thought to belong to $\mathbb{N}$ (specifically, 1), belongs to $\mathbb{R}$.
    
\end{enumerate}

\vspace{.2cm}

Similarly, the preceding sections suggest that similar issues arise with respect to pitch:

\begin{enumerate}[noitemsep,label={(\arabic*)}]
    \item The notion that a unique pitch is an objective feature of a harmonic tone, though seemingly obvious, does not withstand closer examination.
    \item The pitch or pitches perceived from a harmonic tone depend on the listener and the listening conditions.
    \item As shown in Figure~\ref{fig:stamina} (Section~\ref{subsec:fission}),  given a harmonic tone and several listeners, the perceived pitch or pitches form a distribution. The distribution coefficients and the mean number of perceived pitches belong to $\mathbb{R}$, although traditional music theory assumes they belong to $\mathbb{N}$ (typically, a single pitch per tone).
\end{enumerate}

\newpage

Additionally, the coastline paradox illustrates an inductive process (see Section~\ref{sec:methodological}) starting from the observation of particulars and leading to a more general theory (fractals):

\begin{itemize}[noitemsep]

    \item \textbf{Observation:} \citet{steinhaus1954length} noted that irregular forms resist classical measurement, exposing the limits of traditional concepts like length.
    
    \item \textbf{Systematization:} \citet{richardson1961problem} observed that measured boundary lengths increase as scale decreases, suggesting a systematic but unformalised pattern.

    \item \textbf{Model:} \citet{mandelbrot1967long} formalised these findings, introducing fractal dimension and showing natural forms can have non-integer dimensions.
    
\end{itemize}

This chapter suggests that a comparable process may be undertaken in regard to pitch.


\vspace{.4cm}
\section*{Conclusion}

This chapter has suggested that pitch -- often treated as a fixed and objective property of a tone -- may instead be better understood as a perceptual construct, particularly in the case of electronic means of production. Drawing on empirical observations from contemporary popular music (CPM), we propose that even basic traditional assumptions inherited from Western classical music may not hold universally.

\vspace{.2cm}

In particular, the assumptions that a harmonic tone is perceived as a single pitch, and that combining tones with different pitches results in a chord, lead to conflicting outcomes when applied simultaneously. This contradiction is captured by the phenomenon of tonal fusion, in which multiple tones are perceived as a single one. A related concept, tonal fission, describes how a single harmonic tone may give rise to multiple perceived pitches depending on the listener and the listening conditions. This variability can lead to multistable percepts, where the perceived pitch changes over time.

\vspace{.2cm}

In electronic music, tones may precede pitch organisation in the compositional process. In such cases, pitch and tuning systems emerge from the tone's structure, rather than being imposed beforehand. This recalls the historical shift from modal polyphonic writing to harmonic and tonal writing, where harmony evolved from being a resultant effect of counterpoint to an active organising principle. In the present context, pitch may undergo the reverse shift -- from active to resultant.

\vspace{.2cm}

Finally, by drawing a parallel with the coastline paradox, we highlighted how an apparently straightforward measurement -- whether geographic length or perceived pitch -- can unravel under scrutiny. Just as the measured length of a coastline changes with the scale of measurement, the perceived pitch of a tone varies with context and perspective. In both cases, quantities once assumed to belong to $\mathbb{N}$ (natural numbers) may be more accurately described as belonging to $\mathbb{R}$ (real numbers).

\vspace{.2cm}

These findings call for a reevaluation of pitch models across disciplines -- not to discard existing frameworks, but to expand them. By grounding pitch theory in perceptual variability rather than potentially obsolete theoretical idealisation, we move toward a richer understanding of musical experience, especially in contexts that diverge from classical norms.

\newpage
\bibliographystyle{apalike}
\bibliography{mybib.bib}
\end{document}